\documentclass[useAMS,usenatbib]{mn2e}

\usepackage{amsmath,amssymb}
\usepackage[dvips]{graphicx}

\renewcommand{\mathbf}[1]{\boldsymbol{#1}}
\newcommand{\dispra}[3]{$#1^h #2^m #3^s$}
\newcommand{\dispdec}[3]{$#1^\circ #2^\prime #3^{\prime\prime}$}

\title[Large Scale Structure in the ELAIS S1 Survey]{Large Scale Structure in
the ELAIS S1 Survey}

\author[E. A. Gonzalez-Solares et al.]%
{E. A. Gonzalez-Solares$^{1,2}$\thanks{E-mail: eglez@ast.cam.ac.uk},
S. Oliver$^1$, 
C. Gruppioni$^{3,4}$, 
F. Pozzi$^{4,5}$ 
C. Lari$^6$
\newauthor
M. Rowan-Robinson$^7$,
S. Serjeant$^8$,
F. La Franca$^9$,
M. Vaccari$^7$\\
$^1$Astronomy Centre, Department of Physics and Astronomy, University of
Sussex, Falmer, Brighton BN1 9RH \\
$^2$University of Cambridge, Institute of Astronomy, The Observatories, Madingley Road, Cambridge CB3 0HA\\
$^3$Istituto Nazionale di Astrofisica, Osservatorio Astronomico di Padova, 
vicolo dell Osservatorio 5, I-35122 Padova, Italy\\
$^4$Istituto Nazionale di Astrofisica, Osservatorio Astronomico di Bologna, 
via Ranzani 1, I-40127 Bologna, Italy \\
$^5$Dipartimento di Astronomia, Universit`a di Bologna, via Ranzani 1,
I-40127 Bologna, Italy \\
$^6$Istituto di Radioastronomia del CNR, via Gobetti 101, I-40129 Bologna,
Italy \\
$^7$Astrophysics Group, Blackett Laboratory, Imperial College of Science,
Technology \& Medicine, Prince Consort Road, London SW7 2BZ \\
$^8$Centre for Astrophysics and Planetary Science, School of Physical
Sciences, University of Kent, Canterbury, Kent,CT2 7NR \\
$^9$Dipartimento di Astronomia, Universita degli Studi ``Roma TRE'', Via della Vasca Navalle 84, I-00146, Roma, Italy
}

\begin{document}

\date{}

\pagerange{\pageref{firstpage}--\pageref{lastpage}} \pubyear{2003}
\maketitle
\label{firstpage}

\begin{abstract}
  We present an analysis of the two-point angular correlation function
  of the ELAIS S1 survey. The survey covers 4 deg$^2$ and contains 462
  sources detected at 15$\mu$m to a 5$\sigma$ flux limit of 0.45 mJy.
  Using the 329 extragalactic sources not repeated in different
  observations, we detect a significant clustering signal; the
  resulting angular correlation function can be fitted by a
  exponential law $w(\theta) = A \theta^{1-\gamma}$ with $A =
  0.014\pm0.005$ and $\gamma = 2.04\pm0.18$. Assuming a redshift
  distribution of the objects, we invert Limber's equation and deduce
  a spatial correlation length $r_0=4.3^{+0.4}_{-0.7}\,h^{-1}$\,Mpc.
  This is smaller than that obtained from optical surveys but it is in
  agreement with results from IRAS. This extends to higher redshift
  the observational evidence that infrared selected surveys show
  smaller correlation lengths (i.e. reduced clustering amplitudes)
  than optical surveys.
\end{abstract}


\section{Introduction}

Theories of structure formation were strongly constrained by the statistical
measurements of clustering in some of the early galaxy redshift surveys.
Surveys of infrared galaxies, in particular, were able to rule out the then
standard Cold Dark Matter Model \citep{e90, s91}. Present day redshift
surveys such as the 2dFGRS \citep{c01}, SDSS \citep{y00} and, in the
far-infrared, the Point Source Catalog Redshift survey, PSC-z \citep{s00} are
now able to provide definitive measurements of the galaxy clustering in the
local Universe.

Despite this success, we have always known that galaxies are biased tracers
of the matter distribution and yet we have a poor observational or
theoretical understanding of this bias, although it is assumed to be related
to the process of galaxy formation and evolution.  To understand bias and, by
inference, galaxy formation, we need to better understand the clustering of
different galaxy types and the evolution of this clustering with redshift.

In this paper we attempt to provide an estimate of the clustering of infrared
galaxies a factor of ten deeper (in redshift) than those seen in the IRAS
surveys.  To do this we provide the first estimate of clustering from any of
the extragalactic ISO surveys.  This is thus the first estimate of clustering
from galaxies selected at 15$\mu$m.  We have used part of the ELAIS survey
(Oliver et al. 2000) as this probes the largest volume of any of the ISO
surveys. We measure the projected clustering by calculating the angular
correlation function, we then discuss the constraints this places on the three
dimensional clustering using Limber's equation.




\begin{figure*}
  \includegraphics[angle=0,width=0.6\hsize]{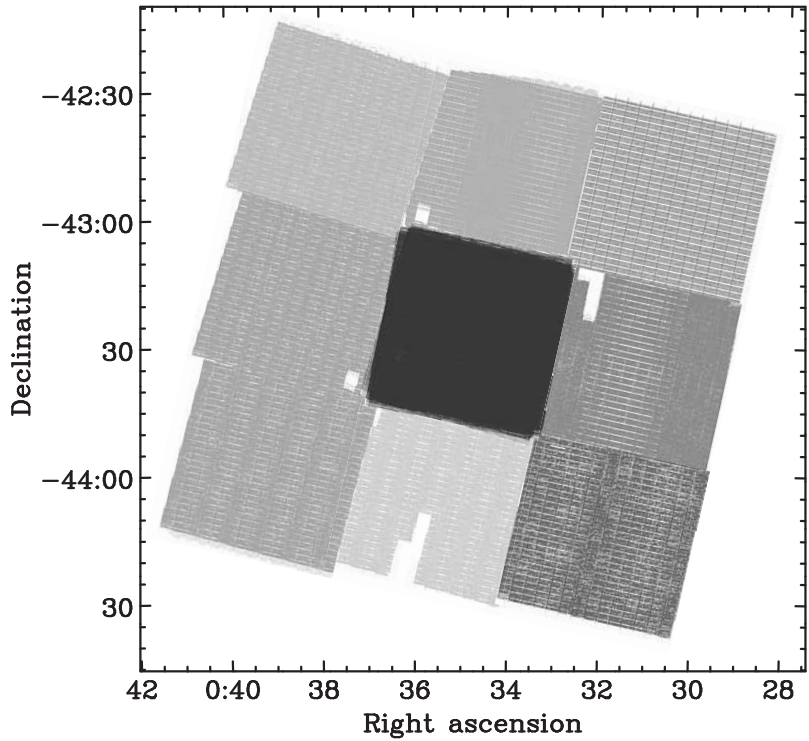}
  \caption{Selection function used to generate the random sample of
    galaxies. The dark central region arises due to the deeper
    observations carried out in.}
  \label{fig:mask}
\end{figure*}

\section{The ELAIS S1 Survey}

The European Large-Area ISO survey (ELAIS, Oliver et al. 2000) was the
largest Open Time programme on ISO.  This project surveyed 12 square degrees,
larger than all but the serendipitous surveys, making it ideal for clustering
studies. The main survey bands were at 6.7, 15, 90 and 170 $\mu$m.  Of these
bands the 15$\mu$m catalogues contain the greatest density of galaxies
\citep[see e.g.][]{g02,se00}, and provide the best
statistics for clustering.  The final analysis of the 15$\mu$m data using the
Lari method  for one of the ELAIS fields (S1) has recently been
completed \citep{l01} and this is the sample that we use in this analysis.

The S1 field is located at $\alpha(2000)~=$~\dispra{00}{34}{44.4},
$\delta(2000)~=$~\dispdec{-43}{28}{12}, covering an area of $2\times 2$
square degrees. The 15\micron\ survey is made from 9 raster observations,
each one of $\sim 40^\prime\times40^\prime$. The central
raster S1\_5 has been  observed three times. Using the Lari method we have
obtained a sample of 462 sources to $5\sigma$ in the flux range
0.45-150\,mJy \citep{g02}. 

\section{Selection function}

Besides the galaxy catalogue itself, the selection function is the most
important ingredient in the calculation of clustering statistics. Errors in
the selection function will invalidate the answer, whereas errors in the
weighting scheme will usually make the answer more noisy.

A selection function is required for each source list that is being
investigated. The selection function, $\phi$, is defined as the expected
number density of sources as a function of $\mathbf{r}$ (which might be two
or three dimensional), in absence of clustering; i.e., the expected number of
galaxies $d\mathcal{N}$ in a volume $dV$ is $d\mathcal{N} = \phi(\mathbf{r})
dV$. With this definition, $\int \phi(\mathbf{r}) dV = \mathcal{N}$. The
selection function is used to simulate a catalogue with no clustering.

To be selected from the ELAIS S1 catalogue sources had to exceed a
signal-to-noise threshold, $\sigma_{min}$. The signal to noise of a
detected source $i$, is $\sigma_i(\mathbf{r_i}) = S_i/N(\mathbf{r_i})$
where $S_i$ is the signal of the source and $N(\mathbf{r_i})$ is the
noise at the position of the source. Had this source been in a
different part of the survey, $\mathbf{r}$ it would have had a
different signal-to-noise.
We can define then a mask, $M_i(\mathbf{r})$,
which represents the detectability of each object as a function of
position as follows
\begin{equation}
M_i(\mathbf{r}) = \left\{
\begin{array}{ll}
0 & {\rm if}\ \sigma_i(\mathbf{r}) < \sigma_{min} \\
1 & {\rm if}\ \sigma_i(\mathbf{r}) \geq \sigma_{min} \\
\end{array}
\right.
\label{eq:mask}
\end{equation}

where $\sigma_{min} = 5$. The un-normalised selection function can
then be written as
\begin{equation}
\phi^\prime = \sum_i M_i(\mathbf{r}) \label{eq:sfunc}
\end{equation}
which can be normalised
\begin{equation}
\phi = \phi^\prime \frac{\mathcal{N}}{\int \phi^\prime dV}
\end{equation}

\subsection{Building the masks}

In the full ELAIS S1 region there are 9 independent noise maps
$N(\mathbf{r})$, corresponding to 9 independent sub-catalogues. Note that
the central noise map is less noisy and the corresponding
sub-catalogue deeper, because the ISO data were already combined
\citep{g02}.  We constructed a selection function as follows: for each
source in the sub-catalogue we calculate the hypothetical
signal-to-noise ratio (defined as the peak flux over the rms value) at
each point in the raster. Where these exceed the extraction
signal-to-noise threshold $\sigma_{min}$ (equation \ref{eq:mask}), the
value of the selection function at that position is incremented
(equation \ref{eq:sfunc}).

The 9 individual selection functions are then combined into a single
one.  Figure~\ref{fig:mask} shows the final image. In the overlap
region only one selection function was used and the final catalogue
excludes sources in that region that arose from the other
sub-catalogues.  Sources with stellar counterparts have also been
removed \citep[see][]{g02} from the catalogue and excluded from the
calculation of the selection function. We end up with a catalogue of
329 sources.

The selection function so obtained is then used to generate the random
catalogues with no clustering, essential to properly calculate the two
point correlation function.

\section{The Angular Correlation Function}

Correlation functions are widely used to study the distribution of sources in
surveys and to derive large scale properties of galaxies. The two-point
spatial correlation function is defined so that
\[ dP = n^2 [ 1 + \xi(r)] dV_1 dV_2 \]
is the joint probability of finding a source in a volume element $dV_1$ and
another source in a volume element $dV_2$. The function $\xi(r)$ is the
excess probability of finding an object compared to a random distribution of
objects.

Similarly, one can define the two point angular correlation function so that
\[ dP = n^2 [ 1 + w(\theta)] d\Omega_1 d\Omega_2 \]
is the joint probability of finding a source in a solid angle element
$d\Omega_1$ and another source in a solid angle element $d\Omega_2$.  These two
statistics are related by Limber's equation (Peebles 1980).

In order to calculate the angular correlation function of mid-IR sources we
use the \citeauthor{ls93} estimator \citep{ls93}
\begin{equation}
w(\theta) = \frac{{\rm [DD]} - 2 {\rm [DR]} + {\rm [RR]}}{\rm [RR]}
\end{equation}
where [DD], [RR] and [DR] represents the normalized number of galaxy-galaxy,
random-random and galaxy-random pairs with angular separation in $(\theta,
\theta+d\theta)$. 

Errors in the calculation of the angular correlation function are
dominated by Poisson noise. The error in each bin can be estimated
using the following expresion \citep{b96b}:
\begin{equation}
\delta w(\theta) = 2 \sqrt{{\frac{1 + w(\theta)}{DD}}}
\end{equation}
where, in this case, $DD$ is the total number of galaxy-galaxy pairs
(not normalized). Errors calculated using this equation are comparable
to the errors obtained from a bootstrap resampling technique~\citep{l86}.

A second source of errors comes from the to finite size of the sample.
In order to correct for this effect we use the random sample to
calculate the integral constraint as (e.g. Infante et al. 1994)
\begin{equation}
\delta = \frac{\sum N_{rr}(\theta)}{\sum N_{rr}(\theta) (1 + w(\theta))}
\end{equation}
and divide the calculated correlation function by this factor, $\delta
= 0.945$.

\begin{figure}
\begin{center}
\includegraphics[width=0.9\hsize]{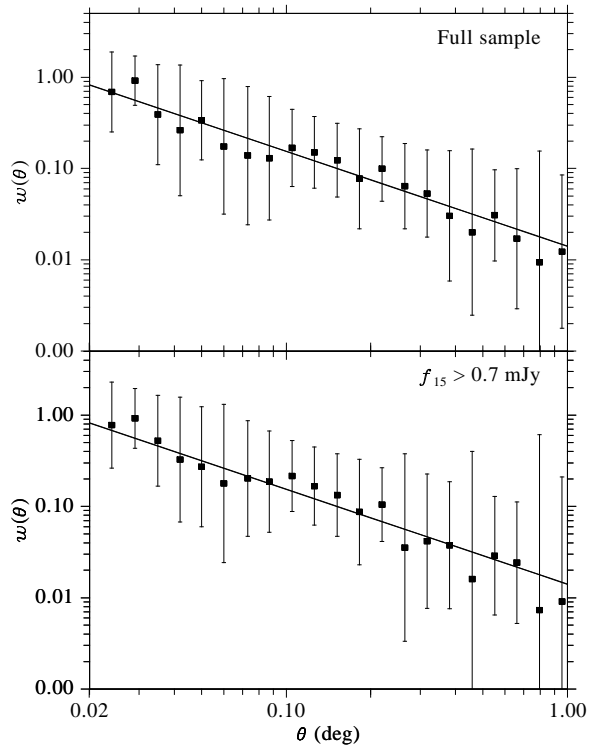}
\caption{Angular correlation function as calculated from ELAIS S1. Top figure 
  shows the calculation performed using all sources, while in the
  bottom figure only those sources with fluxes larger than 0.7 mJy
  have been considered (excluding then, those faint sources only
  detectable in the central deeper raster S1\_5). Data points are
  fitted by a exponential law $w(\theta)=A\theta^{1-\gamma}$ with $A =
  0.014\pm 0.005$ and $\gamma = 2.04\pm 0.18$.}
\label{fig:wtheta}
\end{center}
\end{figure}

Figure \ref{fig:wtheta} (top) shows the obtained angular correlation
function, calculated using 200 realizations with 2000 random sources
each, in intervals of $\log \theta = 0.08$ degrees. Random catalogues
have been built using the selection function calculated in previous
section. Data points have been fitted by an exponential law of the
form $w(\theta)=A\theta^{1-\gamma}$ resulting in $A = 0.014\pm0.005$
and $\gamma = 2.04\pm0.18$ (where $\theta$ is measured in degrees).

The mean number of objects in each field is $\langle n \rangle =
35.25$ (excluding the central raster S1\_5, with 71 sources), with a
standard deviation of 6 objects. Since S1\_5 field reaches deeper flux
limits, it could in principle be subject to clustering variations that
would affect the whole clustering estimation. We repeat the above
calculation removing those sources with fluxes fainter than the flux
limit excluding S1\_5: a total of 27 sources with fluxes lower than
0.7 mJy are removed. By calculating again the selection function and
the angular correlation function we obtain then an estimate of the
clustering from sources detectable over most of the S1 field. In this
case the integral constraint is slightly larger than the previously
obtained, $\delta = 0.970$. The angular correlation function is shown
in figure \ref{fig:wtheta} (bottom) and is fitted by the same
exponential law previously calculated. Although the correlation
function is now slightly larger at scales $\sim 0.1$\,deg, the errors
are also larger and the overall correlation function is well fitted by
the previous function.

\section{Spatial correlation function}




In case of small angles, both $w$ and $\xi$ can be approximated by
power law shapes, and the spatial correlation function can be written
as \citep[e.g., ][]{p78}
\begin{equation}
\xi(r,z) = \left( \frac{r}{r_0} \right)^{-\gamma} (1+z)^{-(3+\epsilon)}
\end{equation}
where $r_0$ is the comoving correlation length at $z=0$ and $r$ the comoving
distance. The parameter $\epsilon$ is the clustering evolution index and is
interpreted as follows. A value $\epsilon=0$ corresponds to stable clustering
in physical coordinates, i.e., galaxy clusters remain unchanged and
clustering changes due to the expansion of the Universe, while
$\epsilon=3-\gamma$ corresponds to clustering fixed in comoving coordinates,
i.e., clustering does not change with time and the galaxy clusters expand
with the Universe.

\begin{figure}
  \begin{center}
  \includegraphics[width=0.9\hsize]{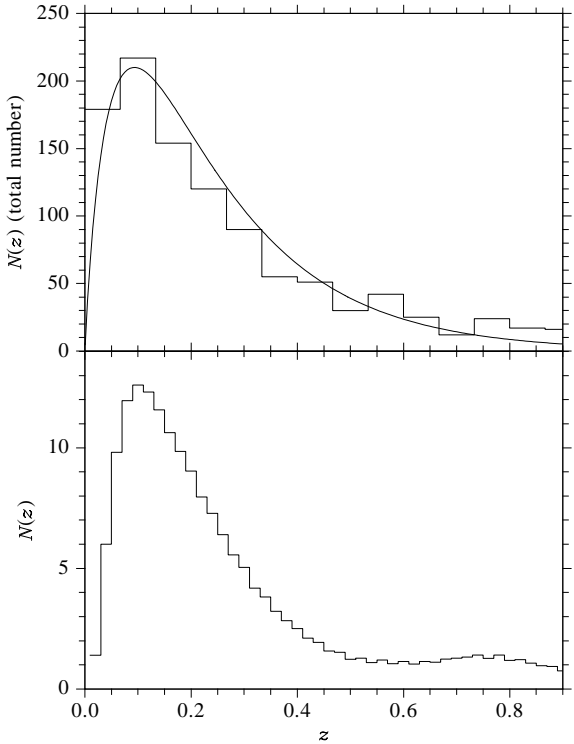}
  \caption{Top: Redshift distribution of ELAIS objects obtained from followup 
    spectroscopic observations and photometric redshifts. Bottom:
    Model redshift distribution of ELAIS sources obtained by
    Pozzi et al. (2004).}
\label{fig:n_z}
\end{center}
\end{figure}

If $w(\theta)$ is parametrized as $w(\theta) = A_w \theta^{1-\gamma}$, then
Limber's equation becomes \citep[e.g.,][]{p78}
\begin{equation}
A_w = C r_0^\gamma \frac{\int D_\theta^{1-\gamma}
  g^{-1}(z)(1+z)^{-(3+\epsilon)} (dN/dz)^2 dz}{(\int (dN/dz) dz)^2}
\label{eq:limber_small}
\end{equation}
where $D_\theta$ is the angular diameter distance, $g(z)$ is the scale factor
multiplied by the element of comoving distance
\begin{equation}
g(z) = \frac{c}{H_0} [(1+z)^2 (1+\Omega_0 z)^{1/2}]^{-1}
\end{equation}
and
\begin{equation}
C=\pi^{1/2} \frac{\Gamma[(\gamma-1)/2]}{\Gamma(\gamma/2)}
\end{equation}
The only unknown quantity in equation~\ref{eq:limber_small} is the
redshift distribution of the sources. A distribution given by
\[
  \frac{d N}{d z} = \frac{3 \mathcal{N} \Omega_s}{2
  z_c^3} z^2 \exp \left[ - \left( \frac{z}{z_c} \right)^{3 / 2} \right]
\]
where $z_m=\sqrt{2} z_c$ is the median redshift of the survey,
provides good fits to the distribution of optical galaxies \citep[and
references therein]{b96} and as been widely used to invert Limber's
equation. However, the redshift distribution of mid-IR sources is more
extended and the equation above is no longer valid.
Figure~\ref{fig:n_z} (top) shows the redshift distribution of
the ELAIS sources, obtained from optical spectroscopy (La Franca et
al, 2004, in prep; Perez-Fournon et al, 2004, in prep) and photometric
redshifts \citep{mrr03}. Instead of the optical $dN/dz$ we use
\begin{equation}
  \frac{d N}{d z} \propto  z \exp \left[ - \left( \frac{z}{z_c} \right)^{3 / 4} \right]
\label{eq:dndz}
\end{equation}
which fits reasonably well the distribution of ELAIS sources.

We can now integrate equation \ref{eq:limber_small} in order to obtain
$r_0$.  Assuming $\Omega_0=1.0$ and $\epsilon=0$ and using a median
redshift of $z_m=0.1$ in equation~\ref{eq:dndz} , we obtain
$r_0=4.3\,h^{-1}$\,Mpc with a 95\% confidence level in the range
$3.8\,h^{-1}$\,Mpc to $4.7\,h^{-1}$\,Mpc.

An alternative redshift distribution of the ELAIS sources has been
presented by \cite{p04}, computed from the luminosity function fit of
galaxies on ELAIS S1 and S2 (see figure~\ref{fig:n_z} -- bottom). When
using this redshift distribution as input to equation
\ref{eq:limber_small} we obtain similar results of
$r_0=4.2\,h^{-1}$\,Mpc with a 95\% confidence level in the range
$3.6\,h^{-1}$\,Mpc to $4.8\,h^{-1}$\,Mpc.

\section{Discussion}

We use a sample of 329 15\micron\ galaxies detected in the ELAIS S1 survey
covering a region of 4 square degrees of sky to determine the angular
correlation function of the galaxies. We measure $w(\theta)$ up to scales of
1 degree, corresponding to $\sim 7$\,Mpc at the mean redshift of $z=0.14$. The
resulting correlation function is well fitted by a power law $w(\theta) = A
\theta^{1-\gamma}$ with $A=0.014\pm0.005$ and $\gamma = 2.04\pm 0.18$.
Assuming a redshift distribution we invert the angular correlation using
Limber's equation in order to determine the correlation length $r_0$. We find
a value of $r_0=4.3\,h^{-1}$\,Mpc with a 95\% confidence level in the range
$3.8\,h^{-1}$\,Mpc to $4.7\,h^{-1}$\,Mpc. 

\begin{table}
\caption{Summary of correlation lengths values obtained from different
  surveys.} \label{tab:r0}
\begin{center}
\begin{tabular}{lcccc}
\hline\hline
\textbf{Survey} & \textbf{$z_m$} & \textbf{$r_0$} & \textbf{$\gamma$} & \textbf{Ref.} \\ \hline
APM                         &  0.05  &  5.7  &  1.67  & 1   \\
SDSS       & 0.18 & 5.7$\pm$0.2 & 1.75$\pm$0.03 &  2 \\
2dFGRS                      &  0.08  & 4.92$\pm$0.27 & 1.71$\pm$0.06 & 3\\
IRAS                       &  0.02  &  3.79$\pm$0.14 &  1.57$\pm$0.03 & 4\\
PSCz                      &   0.02 & 3.7 & 1.69 & 5 \\
ELAIS                     &  0.14  & 4.3$^{+0.4}_{-0.7}$  &  2.04$\pm$0.18 & \\
\hline
\end{tabular}
\end{center}
References: 1 \citealt{m90}; 2 \citealt{z02} (assuming a Einstein-de Sitter
cosmology); 3 \citealt{n01};  4 \citealt{s92}; 5 \citealt{j02}

\end{table}

Table \ref{tab:r0} lists the correlation lengths obtained from several
optical and infrared surveys. While optical surveys show a significant
clustering with $r_0\sim 5\,h^{-1}$\,Mpc, IRAS survey shows a lower
value. The data described in this paper is consistent with this trend:
that mid infrared selected sources show a smaller correlation
length. This is expected because optical surveys favour elliptical
galaxies which are more strongly clustered than spiral galaxies, while
infrared surveys preferentially select spirals and star-forming
galaxies.  We note that the ELAIS selected galaxies appear to have a
marginally steeper two point correlation function than the optical and
the IRAS surveys. By fixing $\gamma$ to the lower value allowed by the
fit $\gamma = 1.86$ we obtain $r_0 = 4.1_{-0.5}^{+0.2}$, bringing the
clustering amplitude closer to that seen by IRAS.

It is interesting to note that the optical correlation function at
$z\sim 0.2$ \citep{z02, n01} is consistent with that at $z \sim0$
\citep{m90}, i.e. there is no apparent evolution in the correlation
function over this albeit small redshift range.  Likewise the infrared
galaxy correlation function estimate from this work at $z\sim 0.14$ is
consistent with the infrared galaxy correlation function estimated
from IRAS galaxies at $z=0$ \citep{j02}. This apparent lack of
evolution may be because evolutionary effects are small or that
evolution in mass clustering is compensated by evolution in galaxy
bias.  It will be interesting to see if this apparent non-evolution in
clustering of different galaxy population mixes is still found in
deeper surveys (e.g. SWIRE, Lonsdale et al. 2003) as this might imply
a striking conspiracy in the evolution of the bias in different galaxy
types.

By performing a systematic analysis of all density peaks in the
redshift distribution of field galaxies, \cite{e04} recently found an
excess of ISO selected galaxies over the whole range of density peaks.
This suggests than infrared galaxies are more strongly clustered than
optical galaxies, in apparent contradiction to our results. Since the
ISO-CAM surveys are deeper than ELAIS, it is plausible that there is
an evolution of the clustering over the redshift interval spanned by
these two ISO surveys. Infrared galaxies would then become more
clustered towards higher redshift, while the clustering of optical
galaxies changes little. This would agree qualitatively with
hierarchical pictures of structure formation. Such models predict that
star formation and galaxy formation is driven by merger rate that
would be a function of environment. Star formation would thus have
occurred first and vigorously in the denser (more clustered) regions
of the Universe, taking more time to initiate in lower density (less
clustered) regions \citep[see also][]{e03}. The strong evolution in
luminosity function of infrared galaxies (e.g. Pozzi et al. 2004)
might then be coupled with an evolution in their clustering. Optically
selected galaxies sample regions where past star formation activity
was high as well as those where current activity is high and are thus
less sensitive to these effects. Finally this apparent contradiction
may simple be that the deep ISO-CAM surveys do not sample a
sufficiently large volume to be representative of the rest of the
Universe.

\section{Acknowledgements}

Eduardo Gonzalez-Solares was supported by PPARC grant
PPA/G/S/2000/00508 and EC Marie Curie Fellowship MCFI-2001-01809.
This paper is based on observations with {\em ISO}, an ESA project, with
instruments funded by ESA Member States (especially the PI countries:
France, Germany, the Netherlands and the United Kingdom) and with
participation of ISAS and NASA.
Some of this work was supported by EEC Training Mobility Research
Network ``Probing the Origin of the Extra-galactic background light''
(POE) HPRN-CT-2000-00138. The author also wants to thank D. Elbaz for
useful comments on the discussion of the results presented.

\label{lastpage}
\end{document}